\documentstyle[eqsecnum,twocolumn,psfig,aps]{revtex}
\draft
\begin{document}
\pagestyle{plain}
\def\be{\begin{equation}}
\def\ee{\end{equation}}
\title{Nonlinear DC-response in Composites: a Percolative Study}
\author{Abhijit Kar Gupta\thanks{abhi@hp1.saha.ernet.in} and 
Asok K. Sen\thanks{asok@hp2.saha.ernet.in}}
\address{LTP Division, Saha Institute of Nuclear Physics\\
1/AF, Bidhannagar, Calcutta 700 064, India}
\date{\today}
\maketitle
\begin{abstract}
The DC-response, namely the $I$-$V$ and $G$-$V$ charateristics, of a variety
of composite materials are in general found to be nonlinear. We attempt to 
understand the generic nature of the response charactersistics and study the 
peculiarities associated with them. Our approach is based on a simple and 
minimal model bond percolative network. We do simulate the resistor network
with appropritate linear and nonlinear bonds and obtain macroscopic 
nonlinear response characteristics. We discuss the associated physics.
An effective medium approximation (EMA) of the corresponding resistor
network is also given.
\end{abstract}


\section{Introduction}           \label{intro}

For measuring the susceptibility of a physical system, one generally studies its
linear response for a small external perturbation such that this perturbation
does not appreciably change the basic nature of the system governing the 
response characteristics under study. For an appropriately large 
perturbation (and a nondestructive one) the physical system may acquire
newer modes of response and hence the system's response
characteristic may change with the externally applied field. 

Composite systems, which are the object of our study here, may be thought to
be comprised of many microscopic elements or grains (much larger than atomic
dimensions) having different physical properties.  In the electrical case,
it means that some of them may be metallic, some insulating and yet some
others semiconducting.  Further, the interface between them may also have
different characteristics.  If one takes the charge carrier (electron or hole) 
as a random walker, a metal corresponds  to  diffusive  motion  giving rise to
Ohm's law. This diffusive behavior changes in the presence  of tunneling
across a barrier, like  $M-I-M$,  $S-M-S$,  $M-s-M$,  $s-I-s$,  etc. junctions
($M$ stands for a metal,  $S$ a superconductor, and $s$ a semiconductor). 
For example, a potential barrier called Schottky barrier arises within 
a semiconductor in intimate contact with a metal and gives rise to current
$i \propto \exp(v/E_0)$, where $E_0$ is function of temperature. In the 
case of a $p$-$n$ junction heavily doped with $Sb$ donors, the current seems to
go from one linear region to another in steps, or the conductance smoothly
connects two plateaus. Qualitatively similar curves \cite{note1} 
are observed for $InP$ tunnel diode, $Au$ doped $Ge$ tunnel diode, {\em etc}. 
Further there could be inelastic processes by which an electron loses eneregy 
to the vibrational modes of species within the barrier. These processes thus 
provide extra paths or channels for current flow (thus increasing the 
conductance), and hence one may observe breaks in $i$-$v$ curves joining two 
piece-wise linear (ohmic) regions.  In the following percolative approach to
obtain the direct current (DC) response of composite systems,
one treats these elements semi-classically in the sense that 
one does not solve  for  the  wave-function  but  takes  the  quantum 
mechanical effects indirectly by including only the tunneling currents 
through barriers \cite{note2}. 

It is clear that if these microscopic tunneling (`nonlinear') elements 
become linear as  discussed  above  beyond  some  characteristic  microscopic 
voltage, then the nonlinear conductance of  the  whole  system  made up of 
such elements would change over  to  ohmic  behavior (constant 
differential  conductance) beyond some system-specific  applied  voltage. 
The point which may not be apparently obvious  is  that  even  if  the 
nonlinear  bonds  remain  so  all  the  way  upto  infinity ({\em i.e.}, 
their resistances become zero (`super'-conductor)  at  infinite  voltage or 
frequency), the macroscopic system comprising of such bonds still saturates at 
high voltage to another linear region or ohmic behavior if the 
tunneling bonds do not percolate by themselves. In either case, {\it the 
resistive behavior of the original percolative structure (zero voltage) 
guides the resistive behavior of the `shorted' percolative structure}. 
Thus, for vanishing driving fields, the system is in 
some state providing a fixed number of channels for the response,  but 
asymptotically for infinitely large driving fields, the system  is  in 
another state representing a  much  higher number of channels.  The 
generalized susceptibility  (which  could  be  thermal  or  electrical 
conductivity, any elastic moduli, or permeability, {\em etc}.) of  this  two 
(or, multi)-level system  crosses  over  from  one  asymptotic  linear 
region (value) to another (of higher saturation  value)  through  some 
nonlinear region. It may be noted here that in rupture type of breakdown
({\em e.g.}, electrical fuse \cite{arcan}) phenomena, essentially the  
opposite  thing happens, namely the system crosses over  from  a  higher 
susceptibility  value (highly efficient and stable network) to a lower one 
where the  system cannot hold itself anymore. In electrical terminology, 
the system as a whole  fuses  to  give  rise  to  an  insulator, or in 
mechanical terminology, the structure becomes mechanically  unstable. 
{\it Thus  this enhancement or  (de-enhancement) of the number of 
system-spanning channels and their eventual saturation as a function of 
some  driving field is at the heart of the nonlinear response} which is 
considered here in this work.

\subsection{Experimental Facts}          \label{facts}

Nonlinear transport characteristics have been revealed in many different types
of materials, including an early work by van Beek and van Pul (1964) on 
carbon-black loaded rubbers. 
References \cite{exp,kkb,rkc,chen} in this regard are meant to be 
representative and not exhaustive.  We discuss below 
some interesting and common features of composite materials, specially which
are highly structured and give rise to some sort of universal behavior. 
Composite materials are typically a mixture of two or more phases, namely
a mixture of metallic kind of material in an insulator where the mixture is
not in the atomic scale. The metallic islands formed in the insulating matrix
(as may be found by tunneling electron micrograph, TEM) are typically of 
dimensions much greater than the atomic size but much smaller than the
macroscopic system size. 
As an example we may here refer to the carbon-black-polyvinylchloride (PVC)
composites \cite{balberg}. Carbon blacks composed of small but complicated 
shaped particles usually exist in the form of ``high-structure'' aggregates, 
whereas smaller and geometrically simpler particle aggregates are also 
possible and they are called ``low-structure'' blacks. A schematic description 
of such composites made of different forms or structures are given in 
Ref.~\cite{balberg}). The conductivity exponents for those three types
of carbon-black-PVC samples were also measured. They were found to be $t=4$, 
$t = 2.8$ and $t = 2$ for ``low-structure'' (commercially called Mogul-L black),
``intermediate-structure'' (Cabot black) and ``high-structure'' composites 
respectively \cite{balberg}. Only the ``high-structure'' composite ($t = 2$) 
was found to be in the universality class of ordinary percolation problems and we
actually have those kind of systems in our mind.  In fact the following 
features pertain to a wide variety of such composite systems available.

\begin{itemize}

\item {{\bf Very Low percolation threshold:} Usually these composites
exhibit an unusually low percolation threshold.  For example, in an experiment 
on carbon-wax system \cite{kkb}, there is a percolation transition at $p_c$ =
0.0076. Very low thresholds are also reported for other systems, 
{\em e.g.}, $p_c=0.002$ for carbon-black-polymer composites \cite{mandal} and 
$p_c=0.003$ for sulphonated (doped) polyaniline networks \cite{reghu}.}

\item{{\bf Qualitatively identical $I$-$V$ (as well as $dI/dV$ vs. $V$)
response both below and above threshold:} As an example one may note the 
experiment on $Ag$ particles in $KCl$ matrix by Chen and Johnson \cite{chen}, 
where similar nonlinear transport was reported both below and  above 
$p_c$ = 0.213, even though the exponents for these two regions were reported to 
be grossly different. The fact that charge is carried even below $p_c$ indicates 
that {\it tunneling/ hopping between disconnected conducting regions} does take 
place and is responsible for the nonlinearity.}

\item{{\bf Power-law  growth  of  conductance:} For small $V$, the excess 
conductance (above the constant/ ohmic part) is claimed to grow as a 
non-integer power law with exponent $\delta$ =  1.36  in  3D  \cite{kkb}, 
whereas previous theoretical works took $\delta$ to be an integer equal to 2 
to keep voltage inversion symmetry. Power-law in 
conductance against voltage ($G$-$V$) automatically implies the power-law in 
$I$-$V$ characteristics. The nonlinearity in $I$-$V$ curves and the associated 
power-law have been observed in a variety of experiments (although in most
of the experiments the $G$-$V$ curves are not examined). 
Rimberg {\em et al.}, \cite{rimberg} measured the $I$-$V$ characteristics
of one- and two-dimensional arrays of normal metal islands connected
by small tunnel junctions and found a power-law: $I \sim (V - V_g)^{\alpha}$, 
$V_g$ being the system's threshold voltage. The nonlinearity exponent $\alpha$
is found to be 1.36 $\pm$ 0.16 in the case of 1D array and 1.80 $\pm$ 0.16
for 2D. However, the values of $\alpha$ were earlier predicted theoretically to
be 1 and 5/3 in 1D and 2D respectively by Middleton {\em et al.}
\cite{middleton}, through an approximate analytical calculation.}

\item{{\bf Saturation of conductance:} In general for composite
systems and for many other materials the $G$-$V$ characteristics show a very 
interesting behavior. The $G$-$V$ curve is seen to saturate for an 
appropriately high enough voltage below the Joule-heating regime. The typical 
curve then looks like a nonlinear sigmoidal type function interpolating two 
linear regimes. Some of the experiments \cite{kkb,chen}
as mentioned above present the $G$-$V$ data in this form for the purpose of
proper analysis. A similar nonlinear curve for conductivity
against field is also presented by Aertsens {\em et al.} \cite{aert} for 
microemulsions of water in oil under an external electric field.} 

\item{{\bf Crossover current for nonlinearity:} The current $I_c$ 
(voltage $V_c$) at which the conductance $G$ differs from the zero-current 
(zero-voltage) conductance $G_0$ by some small ($\sim 1\%$) but arbitrarily 
chosen fraction $\epsilon$ is called the crossover current (voltage). 
It is seen to scale as $I_c \propto G_0^x$ ($V_c \propto G_0^{x-1}$),
where $x$ is called the crossover exponent for nonlinearity.  For carbon-wax
composites, $x \cong$ 1.4 \cite{kkb,rkc} in 3D, and for discontinuous thick 
gold films, $x \cong$ 1.5 as found by Gefen {\em et al.} \cite{exp}. One can 
easily check that this exponent is related to $\delta$ above by 
$\delta = 1/(x-1)$.}

\item{{\bf Temperature-dependent conduction:}
Composite systems display very interesting temperature-dependent conduction 
properties particularly in the low temperature regime where the conduction 
is mainly due to phonon-assisted hopping of the electrons between randomly
spaced localized states. One then needs to consider Mott's variable range 
hopping (VRH) or some of its variations namely log~$G \simeq T^{-\gamma}$ for
low $T$.  The conductance of the sample goes through a maximum as the
temperature is increased still further towards some type of metallic behavior.}
\end{itemize}

The first thing to note from the enlisted facts is the ultra-low percolation 
threshold even when the included conducting phases are isotropic. 
This alongwith the fact that many of these
nonlinear systems carry current even below $p_c$ indicates strongly
that tunneling through disconnected (dispersed) metallic regions
must give some virtually connected percolating clusters.  
>From the nonlinear $I$-$V$ characteristics ({\em e.g.}, see the experiment
by Chen and Johnson \cite{chen}) it is observed that the response behavior is 
reversible with respect to the applied field in the sense that the response
curve (current or voltage) does almost trace back as the field is decreased.
This fact also indicates that reversible tunneling is responsible for 
such a behavior. Also the temperature-dependent conductance with a maximum at
some characteristic temperature (dependent on the amount of disorder
present) and the Mott VRH type behavior at very low temperatures
give further credence to tunneling assisted percolation.
So, in the following section, we propose a semi-classical (or, semi-quantum)
model of percolation \cite{skg} which works on
the borderline between a classical and a quantum picture.  Quantum
physics enters our discussion through the possibility of tunneling
of a charge carrier through a barrier (which does not exist classically).

One may consider a wide variety of nonlinear transport (apart from the
nonlinear electrical conduction) processes and study them under the framework 
of percolation theory. One also exploits the key features and ideas of one 
such model and incorporate them into the other. As an example, one
draws analogy between laminar flow in tubes and the electrical currents
and uses the language of electrical network for convenience. The
volumetric flow rate $q$ is identified with the current $i$ and the 
pressure drop $\Delta P$ across a tube or a pore to the voltage difference
$v$ across a bond. Thus, for example, the flow of polymers is modelled
(see {\em e.g.,} Ref.~\cite{sahimi1} and references therein) considering 
a power law $i = gv^{\alpha}$ to each bond (tube or pore), where $g$
is a generalized conductance. Considering the percolating network model
of porous media one employs Monte-Carlo simulations or an effective medium
approximation to calculate
the rheological properties of a power-law fluid in flow through porous 
media. Foam is a non-Newtonian fluid which is used in displacement and 
enhancement of oil recovery from the porous rocks. However, to move the foams 
through the pores, external pressure has to exceed a certain critical value 
$P_c$. In brittle fracture, no microcrack nucleation process takes place 
unless the applied stress exceeds a critical value for the system. A convenient 
description of this particular problem is done through the electrical 
analogoue of breaking with the introduction of random fuse model \cite{arcan}.
One defines the `fuse' as a device with a  constant conductance when the 
applied voltage across it is less than a certain critical value $v_c$, beyond
which it is an insulator. This electrical network model is a scalar analogue of 
the corresponding vector elasticity problem where the former is simpler 
to handle.

In most of the works we mentioned above, dilution plays an important role.
These are models where the percolation theory has been the underlying
framework and many of the interesting properties may be related to the
cluster statistics.  In this paper we first discuss briefly our proposed
model for studying the effective nonlinear conduction properties of various
composites.  We then present the nonlinear current-voltage ($I$-$V$)
characteristics in our model system and discuss the associated details.

\section{The Model and its Percolative Aspects}           \label{model}

To mimic charge transport in composite systems or dispersed metals, we
assume the grains (metallic or metal-like) to be much larger than atoms
but much smaller than laboratory-scale macroscopic objects.  These grains
are randomly placed in the host material which are insulators.
Further we assume quantum mechanical {\it tunneling}
between such grains across some potential barriers.
Clearly these barriers would depend on the local geometry of the insulating
and the metallic grains.
Since in practice, the tunneling conductance should fall off exponentially,
the tunneling should have some length scale 
designating an upper cut-off (for tunneling to occur) in 
the separation between two metallic grains.  Further, the separation between
two grains or the potential barrier can vary continuously
between zero and some upper cut-off. For simplicity and to capture the 
basic physics, we construct a bond (lattice) percolation model for
this problem, such that tunneling may take place only between two
nearest neighbor ohmic conductors (or, $o$-bonds) and no further. Thus one 
may imagine a virtual bond sitting at each such gap which conducts current 
nonlinearly due to tunneling phenomenon.  We call these tunneling 
conductors as tunneling bonds (or, $t$-bonds).

\begin{figure}
\psfig{figure=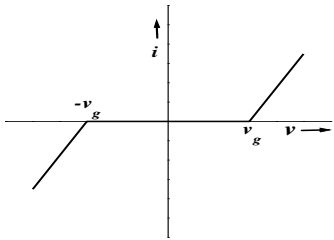,width=3cm,height=4cm}
\caption{The piecewise linear current-voltage ($i$-$v$) characteristic of 
each tunneling conductor.}
\label{fig:ivtun}
\end{figure}

We stress again that our approach would be to solve an appropriate electrical
network based on a semi-classical (or, semi-quantum) percolation model.  
Made of both random resistive and tunneling elements, this network would be
called a random resistor cum tunneling-bond network (RRTN) \cite{drrn}.
Now tunneling may take place through the tunneling bonds in various ways, so 
that the functional form of the tunneling current as a {\it nonlinear} 
function of the potential difference across them may be quite complicated. For 
simplicity, we address the aspects of nonlinearity in a macroscopic system 
which comes through two piecewise linear regions of a $t$-bond such that the
conductance is zero upto a threshold voltage and a given non-zero value beyond.
In this context, it may be noted that disorder in many quantum systems like
charge-density-wave (CDW) systems or flux-vortex lattices of type-II 
superconductors can give rise to `pinning' or inhibition to transport upto
a critical value of the applied field above which tunneling is active. 
The piecewise linear transport is in fact a highly nonlinear process as there 
is a cusp singularity at the intersection point. The transport due to 
tunneling which is the source of nonlinearity in the experimental systems 
\cite{rkc,chen} we focus on, can be well approximated in this way and thus the 
nonlinearity of the macroscopic systems may be understood at a qualitative 
(and, sometimes even at a quantitative) level. Next, one notes as discussed 
above that in many physical systems, the response is negligibly low (or there 
is no response at all) until and unless the driving force exceeds a certain 
threshold value. A class of problems exist where sharp thresholds to 
transport occur. The examples in the electrical case are a Zener diode, a 
CDW system or a type-II superconductor and in the fluid permeability problems, 
for example, a Bingham fluid (where there is a critical shear stress 
$\tau_c$, above which it has a finite viscosity and below which it is so 
enormously viscous that it does not flow). In our RRTN model, we work with 
$t$-bonds which have zero conductance below a threshold; see
Fig.~\ref{fig:ivtun}.

Let us look at the {\it uncorrelated} random bond percolation first.  In this
model, the conducting $o$-bonds are occupied randomly with a certain volume
fraction $p$. One uses a random number generator which generates random numbers 
distributed uniformly in the interval [0, 1], to fix the positions of $o$-bonds
in a certain configuration. So the system now has $p$ fraction of occupied 
bonds and $q$ ($\equiv$ 1-$p$) fraction of unoccupied bonds. 
Upto this it is the standard (uncorrelated) {\it bond percolation} problem
where the percolation threshold in a square lattice is $p_c = 0.5$.
A geometrical connection (or a system-spanning cluster of $o$-bonds) is 
established from one end of the `configuration averaged' system to the other 
only at or above this volume fraction in an infinite size system. If the 
occupied bonds are ohmic conductors and the unoccupied bonds correspond to
insulators (the host) then we have the corresponding random resistor
network (RRN) which carries an average non-zero current only when the
volume fraction, $p > p_c$. Clearly this response is linear or ohmic.

\begin{figure}
\psfig{figure=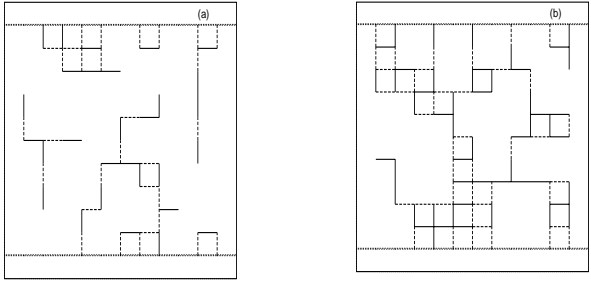,width=6.5cm,height=4cm}
\caption{Tow typical configurations of a square lattice of size $10\times10$ 
with $p < p_c$ and an arbitrary but very large voltage where all the
tunneling bonds indicated by broken lines are active. ($a$)~ For $p$ = 0.15 
the system is seen to have no connecting path across its two ends even when 
one considers the tunneling bonds. ($b$)~This configuration at $p$ = 0.25 is 
seen to have no connecting path with only ohmic bonds, but has at least one 
such path comprised of both the ohmic and the tunneling bonds.}
\label{fig:lattice}
\end{figure}

But, our percolative model is not just a random mixture of two phases.  For
our convenience we take a square lattice in 2D. The basic physics should
remain the same if we go over to 3D.  As usual, the $o$-bonds  are
thrown at random at a certain volume fraction $p$.  The rest $(1-p)$ fraction
contains insulators.  Now we allow tunneling bonds ($t$-bonds) only  across
the nearest-neighbor ($nn$) gaps of two conducting bonds (and no
further) if an appropriate voltage is
applied externally across two opposite sides of the RRTN.
The RRTN model is now comprised of ohmic 
conductors, pure insulators and some non-ohmic (tunneling) conductors. 
To have a clearer view of the proposed model system we refer to the 
Figs.~\ref{fig:lattice} where we show all the ohmic bonds thrown at a 
concentration $p$, and all possible $t$-bonds. 
In the Fig.~\ref{fig:lattice}($a$), we show a realization of a square 
lattice at $p$ = 0.15 in 2D where the system does not have any connectivity 
from one end to the other even if we consider all the $t$-bonds to be active. 
In the Fig.~\ref{fig:lattice}($b$) at $p$ = 0.25,
there is a system spanning cluster (network) of $o$-bonds and $t$-bonds even
though the system is actually below the geometrical threshold ($p < p_c$). 
This demonstrates that the effective percolation threshold \cite{note3} of the 
system is lowered as the $t$-bonds come into play. Thus even if the system is
below $p_c$, and hence has zero conductivity at a vanishingly small voltage,
its conductivity will be non-zero and keep growing at appropriately higher 
voltages, and hence the system will behave non-ohmically. If the system
is already above the geometrical threshold ($p_c$ = 0.5) then the $t$-bonds
simply add extra paths to the already existing current carrying network 
(backbone) of $o$-bonds and give rise to similar non-ohmic behavior.

If we consider the percolative aspects of the model system at very large 
voltages (suppose all the $t$-bonds overcome their threshold and conduct 
ohmically), then this is no more a pure bond percolation problem of a random 
binary mixture. Rather we may think of it as a very specific {\it correlated 
bond percolation} problem. This is because the positions of the $t$-bonds are 
totally correlated to the positions of randomly thrown $o$-bonds. The disorder 
is in the position of the $o$-bonds only. Once the positions of the $o$-bonds 
are given for a particular configuration, the positions of the $t$-bonds are 
automatically determined for that configuration in our model.
We have found out earlier \cite{kgs} that the new percolation threshold with
all the $t$-bonds added is $p_{ct} \approx 0.181$.
We have also addressed \cite{kgs} the question of universality class 
of this problem in this limit and found that the problem belongs to the same 
universality class as that of the pure (uncorrelated) bond or site 
percolation. 

\section{Effective Medium Approximation (EMA) for the 
Correlated Bond Percolation Model}        \label{ema}

An effective medium approximation gives reasonable result for any $p$ away
from the threshold (here $p_{ct}$). The EMA has been used to 
calculate fairly accurate conductivity behavior for general binary mixtures
except in the vicinity of the critical regions. This approach is old and 
had been devised for the transport properties of inhomogeneous materials
first by Bruggeman \cite{brugg} and then independently by Landauer 
\cite{land}. 
Its successful application to the percolation theory by 
Kirkpatrick \cite{kirk} has drawn attention of many others in this field. 
It has then been applied for a wide variety of inhomogeneous materials. 
The development has some similarity with the well-known coherent potential 
approximation (CPA) for treating the electronic properties of the binary 
alloy problem.
The composite systems have seen a series of attempts in this direction while
dealing with the nonlinearity in the response involved (see Ref.~\cite{hui}
and references therein).

Here we follow the method as described in ref.~\cite{kirk}.
The method is as follows.
Consider a random electrical network on a hypercubic lattice of dimension
$d > 1$ ($d=2$ for 2D, $d=3$ for 3D {\em etc.}). The basic idea is to replace a 
random network by a homogeneous effective network or an effective medium 
where each bond has the same average or effective
conductivity $G_e$. The value of the unknown $G_e$ is calculated in a 
self-consistent manner. To accomplish this, one bond embedded in the effective
medium is assigned the
conductivity distribution of the actual random network. The value of $G_e$ 
is then determined with the condition that the voltage fluctuation across the
special bond within the effective medium, when averaged over the proper 
conductivity distribution, is zero.
The voltage ($v$) developed across such a special bond can be calculated 
\cite{kirk} for a discrete lattice of $z$ (= 2$d$ for a hypercubic lattice)
nearest neighbors as
\be
v \propto (G_e - g)/[g + (d - 1)G_e].
\ee
The requirement is that the average of $v$ be zero when the conductance for the
special bond may take any of the ohmic, the tunneling or the insulating bond 
value with appropriate probabilities for the  actual network.

The probability of a bond to be ohmic, tunneling or purely insulating 
according to the considerations of our model is given below for 2D square lattice.
\begin{eqnarray}
P_o & = & p,\\
P_t & = & (p^3+3p^2q+3pq^2)^2q,\\
P_i & = & [1-(p^3+3p^2q+3pq^2)^2]q,
\end{eqnarray}
where $q=1-p$.
For a distribution of three types of resistors, we have
\be
f(g) = P_o\delta(g - g_o) + P_t\delta(g - g_t) + P_i\delta(g - g_i), \label{prob}
\ee
where $g_o$, $g_t$ and $g_i$ are the conductances of ohmic, tunneling and the 
insulating resistors respectively.
The EMA condition stated above {\em i.e.}, $<v>$ = 0 reads
\be
\int dgf(g)(G_e - g)/[g + (d-1)G_e] = 0. \label{int}
\ee
Putting Eq.~(\ref{prob}) into the above Eq.~(\ref{int}), we get 
\be
{P_o(G_e - g_o) \over [g_o + (d-1)G_e]} + {P_t(G_e - g_t) 
\over [g_t + (d-1)G_e]} + {P_i(G_e - g_i) \over [g_i + 
(d-1)G_e]} = 0.   \label{emaeq} 
\ee
The above equation reduces to the quadratic equation
\be
AG_e^2 + BG_e + C = 0,  \label{quad}
\ee
where $A = (d-1)^2$, $B = [(d-1)(P_t + P_i)g_o + (d-1)(P_o + 
P_i)g_t - (d-1)^2(P_og_o + P_tg_t)]$ and 
$C = [P_ig_og_t - (d-1)g_og_t(P_o + P_t)]$,
considering the conductance of the insulating bonds to be $g_i$ = 0.
The valid solution of Eq.~(\ref{quad}) is
\be
G_e = {- B + (B^2 - 4AC)^{1/2} \over 2A}.  \label{effcon}
\ee
Now one can obtain the linear conductance of the macroscopic model composite
system in 2D and 3D, putting d=2 and 3 respectively, given some
specific values or functional forms for the conductances of the elementary
components like the ohmic and the tunneling bonds.

In our model we have assumed that the conductance of a tunneling bond, when it
overcomes its voltage threshold, is the same as that of an ohmic bond.  We believe
that this assumption does not change the phase transition characteristics.  Now in 
the limit of large enough external voltage if we assume $g_t = g_o = 1$ and 
$g_i$ = 0, we have 
\be
G_e = {d(P_o + P_t) - 1 \over (d-1)}.  \label{sol}
\ee
At the percolation threshold $p_{ct}$ the 
effective conductance ($G_e$) of the system  is zero. In 2D 
we now have the following equation putting $d = 2$ in Eq.~(\ref{sol}),
\be
2(P_o + P_t) = 1. \label{emapct}
\ee
Solution of the above equation gives $p = p_{ct}$ =1/4.
This may be compared with our simulation result \cite{kgs} which gives $p_{ct} 
\cong 0.181$. The EMA is basically a 
mean field calculation which overestimates the percolation threshold value
in lower dimensions. However, for pure bond percolation in 2D, this gives
the exact result $p_c = 1/2$. This is so because the square lattice in 2D is 
self-dual in the case of purely random bond percolation problem.

We may also find the value of $p_{ct}$ for 3D where the probability of 
tunneling bond is
\be
P_t = (p^5 + 5p^4q + 10p^3q^2 + 10p^2q^3 + 5pq^4)^2q
\ee
Using $d$ = 3 in Eq.~(\ref{sol}), an equation similar to Eq.~(\ref{emapct})
is obtained, which when solved gives $p_{ct} = 1/8$ exactly.

\section{The Nonlinear $I$-$V$ Characteristics}  \label{iv}

For the work presented below, we make the simplifying assumption that all the
tunneling bonds ($t$-bonds) in our RRTN model have an identical voltage
threshold ($v_g$) below which they are perfect insulators and above which
they behave as the ohmic bonds.  One could certainly introduce disorder by
making $v_g$ random as in Ref. \cite{rh}.  In our case disorder is already
introduced through random positioning of the bonds, and we believe that
our assumption should not affect the dilution-induced nonlinearity exponents.
Indeed, as shown in the sequel, our model gives richer possibilities (dilution
dependence) for the nonlinearity exponent in a composite system.

\begin{figure}
\psfig{figure=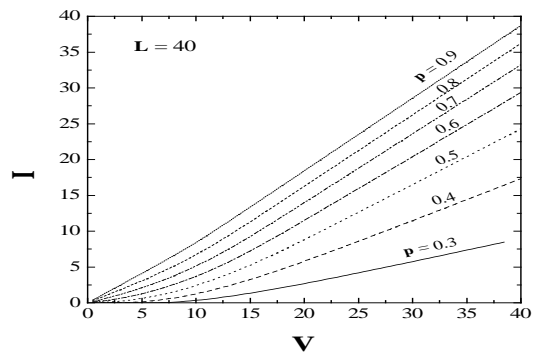,width=4cm,height=6cm}
\caption{A set of nonlinear $I$-$V$ curves for $p$ = 0.3 to 0.9 for the 
RRTN network of size $40\times40$. The current response is averaged over 50
configurations each.}
\label{fig:ivset}
\end{figure}

As the externally applied field (DC) is increased beyond some macroscopic
threshold, some of the $t$-bonds overcome their microscopic thresholds and
may thus increase the overall conductance of the system if the process leads
to newer parallel connectivities for the whole macroscopic composite. Our
computer simulation to study these effects \cite{thesis} involves the 
solution of Kirchhoff's law of current conservation at the nodes of the RRTN
with the linear and the nonlinear (assumed piecewise linear) resistors 
and the standard Gauss-Seidel relaxation.  Current response was averaged over
50 configurations in all the cases mentioned here.
We obtain current ($I$) against voltage $V$ and therefrom the differential 
conductance ($G = dI/dV$) for the whole network at a given volume fraction 
$p$ of the ohmic bonds.
Simulation results for nonlinear $I$-$V$ curves for a square lattice of size 
$L=40$ are plotted in Fig.~\ref{fig:ivset} for $p$ = 0.3 to 0.9. 
For $p > p_c$, one may note that the $I$-$V$ curve is linear up to a certain
voltage ($V_g$), beyond which the nonlinearity shows up.  For $p < p_c$, there
is no current (zero conductance) below a threshold voltage ($V_g$), beyond
which the nonlinear conduction starts.  Nonlinearity is always there in the
$I$-$V$ response for any value of $p$ in the interval $p_{ct} < p < 1$. 
However, for $p_{ct} < p < p_c$, there is no system-spanning path with the 
ohmic bonds (in an average sense) and the response (average current) is zero 
for small voltages. The response in this case starts out nonlinearly from a
non-zero average threshold voltage.
On the other hand, for $p > p_c$, the system always has a conducting path 
through the ohmic bonds (again on an average), and so up to a certain voltage 
($V_g$) there is a constant non-zero conductance and the average response is 
linear.  As $V$ is increased beyond $V_g$, more and more current carrying
paths are added with the help of the $t$-bonds and in effect the conductance
starts increasing with $V$.

\begin{figure}
\psfig{figure=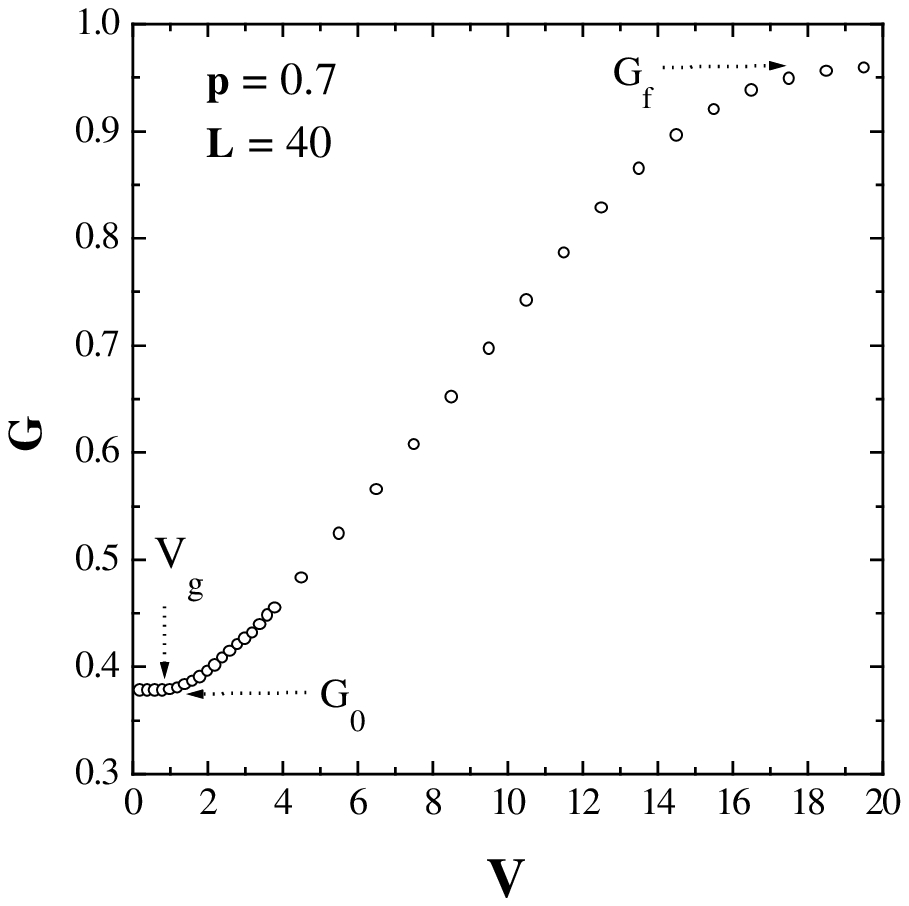,width=4cm,height=6cm}
\caption{The $G$-$V$ curve corresponding to $p$ =  0.7 of Fig.~\ref{fig:ivset}
($p > p_c$). 
The threshold voltage ($V_g$) is indicated in the Figure. Initial conductance 
($G_0 \ne 0$) and the final conductance ($G_f$) are also indicated.} 
\label{fig:gvp7}
\end{figure}

\begin{figure}
\psfig{figure=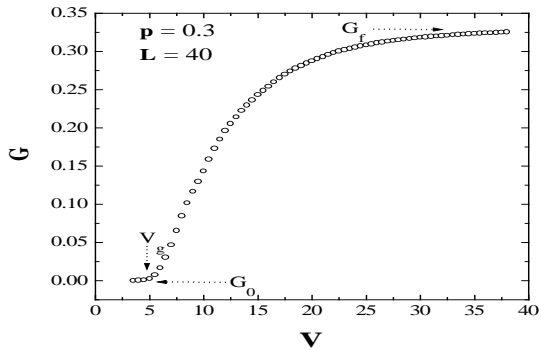,width=4cm,height=6cm}
\caption{The $G$-$V$ curve corresponding to $p$ = 0.3 of Fig.~\ref{fig:ivset}
($p < p_c$). 
The threshold voltage ($V_g$) is indicated in the Figure. Initial conductance 
$G_0 = 0$ and the final conductance ($G_f$) is as indicated.} 
\label{fig:gvp3}
\end{figure}

We find that the entire nonlinear regime of a $I$-$V$ curve can not be 
fitted by a simple power-law which in general may be fitted by a polynomial
function. Even after doing that, the {\it exponent of nonlinearity} from the
fitting of the various $I$-$V$ curves remained somewhat ambiguous since the 
fitting was not very robust. Thus to have a better idea, we obtain the 
differential conductance ($G = dI/dV$) by taking the numerical derivative of 
the the $I$-$V$ data.  In these derived $G$-$V$ curves, one can identify the 
threshold voltage $V_g$ as the onset of nonlinearity and the onset of the
saturation regime ($G_f$) 
much more clearly than in the $I$-$V$ curves. The $G$-$V$ curves corresponding
to $p=$ 0.7 and 0.3 in the Figure~\ref{fig:ivset} are shown in the 
Figs.~\ref{fig:gvp7} and \ref{fig:gvp3} respectively. 
As expected for the two different cases, in 
Fig.~\ref{fig:gvp7} ($p > p_c$) the initial conductance has a finite 
nonzero value ($G_0 \ne 0$), while in the Fig.~\ref{fig:gvp3} 
($p < p_c$) the initial conductance is zero ($G_0 = 0$).  
The general nature of the $G$-$V$ characteristics is that eventually all of
them become flat, {\em i.e.}, assume a configuration- and $p$-dependent
constant conductance beyond some very large voltage $V_s$.
The reason is that for a {\it finite-sized} system there is always a very
large but {\it finite voltage} $V_s$ at which the conductance
of the whole system saturates to an upper maximum value $G_f$ (another linear
regime) since all the possible $t$-bonds become activated and no more channel
parallel to the backbone come into play for a further increment of $V$.
Below we concentrate on the analysis of the
nonlinear conductance behavior of the model system in an effort to first find
a general functional form to describe all these different $G$-$V$ curves and 
then to find the exponent of nonlinearity at different volume fractions.

\subsection{Analysis of the Behaviour of Conductance}    \label{gvana}

\begin{figure}
\psfig{figure=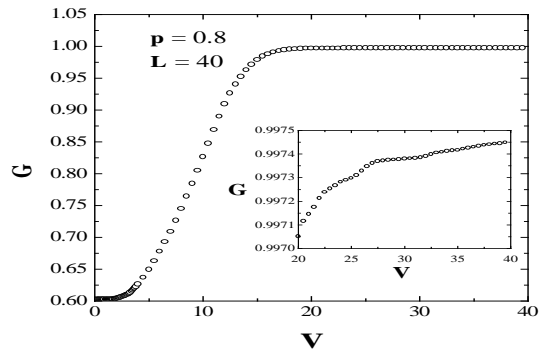,width=4cm,height=6cm}
\caption{A typical $G$-$V$ curve for $L$ = 40 and $p$ = 0.8 with an apparent 
onset of the saturation regime beyond $V \approx$ 20. The inset shows that the
$G$ does not yet saturate in the true sense and instead ever increases in the
regime $20 < V < 40$.}
\label{fig:gvsat}
\end{figure}

Guided by the conductance behavior of a complex system made out of many simple
prototype circuits \cite{skg} and by the fact that the initial power-law
law type growth of $G$ beyond $G_0$ finally saturates to $G_f$, we try to fit
the whole region in our simulation by a function of the form:
\begin{eqnarray}
  G & = & G_0~~~~~~~~~~~~~~~~~~~~~~~~~~~~~~~~~~\mbox{($V < V_g$)}\\
& = & G_f - G_d[1 + \lambda \Delta V^{\mu}]^{-\gamma},~~~~~~~\mbox{($V \ge V_g$)}, 
\label{pfit} 
\end{eqnarray}
where $\Delta V = V - V_g$ is the shifted voltage from where 
nonlinearity appears. For concreteness, we discuss here the fitting of a sample
data set for $L = 40$, $p = 0.6$. For this sample $G_0 \cong 0.154$, $G_f \cong
0.881$. 
The parameter for the best fit in this case as obtained by a simplex search
procedure are $\lambda \cong 3.27\times10^{-4}$, $\mu \cong 1.408$, and 
$\gamma = 125$. Such large 
values of $\gamma$ are obtained for all the cases studied and yet the approach 
to $G_f$ was slower than in actual data. {\it This obviously indicates that 
the approach to saturation is not a power-law type and probably an exponential
function is involved.} At this point we refer to
the Fig.~\ref{fig:gvsat} where we have shown a $G$-$V$ curve for $p = 0.8$ 
and $L=40$ for which the conductance ($G$) seems to saturate (to the naked eye) 
at a voltage above $V$ = 20 and one can see a practically flat regime in 
between $V$ = 20 to 40.
However, in the same figure we have demonstrated in the inset by zooming in
on the y-axis that the actual saturation is yet to come and that the 
conductance ($G$) in this regime increases very slowly with $V$. The reason is 
that there are some tunneling bonds (typically in the transverse direction to 
the electric field) which do not become active even with the application of a 
very large voltage implying that the conductance for the whole
system is yet to reach the complete saturation. This pathology supports the 
fact that the final saturation occurs at an extremely large voltage, and that 
the approach to the saturation is indeed very slow.

\begin{figure}
\psfig{figure=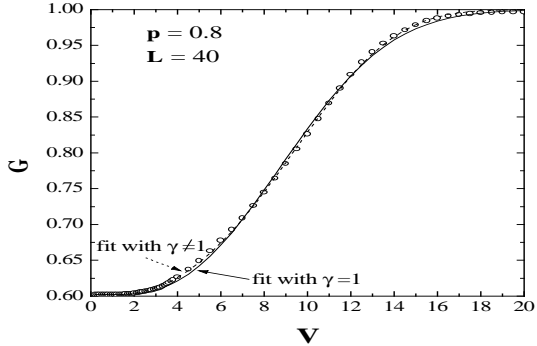,width=4cm,height=6cm}
\caption{The demonstration of the fitting of a $G$-$V$ curve by the proposed 
function (\ref{gfit}) shown for $L$ = 40 and $p$ = 0.8. The fitting is better 
with $\gamma \ne 1$ (shown in dashed line) for the entire data set than that 
with $\gamma = 1$ (shown in full line).}
\label{fig:gvfit}
\end{figure}

We tried the following four plausible functions involving exponentials for the 
entire nonlinear regime replacing the above Eq.~(\ref{pfit}).
\begin{eqnarray}
G & = & G_0 + G_d \exp(-\lambda/\Delta V)\\                              
& = & G_0 + G_d \tanh(\lambda\Delta V^{\alpha})\\
& = & G_f - G_d \exp[1 - \exp(\lambda\Delta V^{\alpha})]\\
& = & G_0 + G_d [1 - \exp(-\lambda\Delta V^{\mu})]^{\gamma}    \label{exfit}               
\end{eqnarray}
Out of all the optimally fitted functions as described above, the last  
Eq.~(\ref{exfit}) is the best in the sense that it gives the minimum mean 
square deviations (MSD) for the $G$-$V$ data.
It may be noted that a special case of Eq.~(\ref{exfit}) with $\gamma$ = 1 
was the form of nonlinearity used by Chen and Johnson \cite{chen} in fitting 
their experimental conductance against voltage data for a composite system 
of $Ag$-$KCl$. But for all $p$ and $L$'s considered by us, $\gamma$ = 1 was 
found inadequate for fitting the typical sigmoidal curves. For example we have 
shown in Fig.~\ref{fig:gvfit}, the fitting with the function 
(\ref{exfit}) for $p$ = 0.8, $L$ = 40: the restricted case of $\gamma = 1$
shown by full line, and an unrestricted, optimally fit $\gamma \ne $ 1 by 
dashed line. Clearly the unrestricted case fits the data extremely well and 
gives an MSD which is much smaller than that of the restricted case as
mentioned above. 

\subsection{The Nonlinearity and the Crossover Exponent}    \label{nonlinexpo}

The conductance ($G$) for metal-insulator composite systems starts growing 
nonlinearly with the applied voltage $V$ from an 
initial value ($G_0$) and finally saturates to a value $G_f$ (as indicated
in the Figures \ref{fig:gvp7} and \ref{fig:gvp3}.
$G_0$ is the conductance when none of the $t$-bonds is active; this happens in
the usual percolation model, without tunneling.  Likewise, $G_f$ is the
conductance when all the so-called tunneling bonds ($t$-bonds)
are actually active and taking part in the conduction.
There are three distinct regimes which, in general, can be precisely located
from the $G$-$V$ characteristics than from the $I$-$V$ charactersistics:

\begin{itemize}

\item (i)~Upto some voltage $V_g$  the initial conductance ($G_0$)
of the system is either zero or a fixed finite value depending on whether the 
system has initially conducting path through the ohmic (or metallic) bonds or
not (see Figs.~\ref{fig:gvp7} and \ref{fig:gvp3}).

\item (ii)~Beyond $V_g$ the nonlinearity starts showing up.
The conductance ($G$) is nonlinear in the regime $V_g < V < V_s$.

\item (iii)~Beyond the voltage $V_s$, $G$ saturates to a voltage $G_f$
(see Figs.~\ref{fig:gvp7} and \ref{fig:gvp3}).

\end{itemize}
The data obtained through our model system in 2D were fitted through the
following general formula as discussed above:
\be
G = G_0 + G_d [1 - {\rm exp}(-\lambda\Delta V^{\mu})]^{\gamma},    \label{gfit}
\ee
where $G_d = G_f - G_0$.  Clearly, irrespective of the value of $V_g$, 
$G_0$ is the conductance in the limit $V \rightarrow 0$.  Experimentally $G_f$
may be obtained by applying a large enough voltage such that Joule heating
remains unimportant. In our computer simulation on finite sized systems, 
we find $V_s$ to be a large but finite voltage (in arbitrary units). 
For example for squares with $L$=40, $V_s$ is found to be typically of the
order of $10^4$-$10^6$.

In a recent experiment by Chen and Johnson \cite{chen} in $Ag$-$KCl$ composite 
(silver particles in $KCl$ matrix), very similar $G$-$V$ curves were 
obtained and the non-ohmic effect was postulated to arise from a localized 
reversible dielectric breakdown between narrowly separated metal clusters
in the metal-insulator composite. The intercluster or interparticle spacing
may have some distribution which is related to the fractal dimension $d_f$ of
the network at the threshold. Chen and Johnson \cite {chen} used the following 
conductance behavior for their data:
\be
G = G_0 + (G_f - G_0)[1 - \exp(-V/V_g)^{n(d_f)}],            \label{chej}
\ee
which is a special form of the function (\ref{gfit}) we propose.  We find 
Eq.~(\ref{chej}) to
be inadequate for the representation of our and experimental data (see above).
The exponent $n(d_f)$ is in fact the same as $\delta$ and is hence
related to the {\it nonlinearity exponent} $\alpha$ (see Sec.~\ref{facts}) by
$\alpha = n(d_f) + 1$.  It has been further shown in the above work that 
$n(d_f)$ increases as the silver volume fraction is decreased and it shows a 
sharp change at the threshold. We shall discuss in this section how our
model captures this dilution-dependent nonlinearity exponent.

\begin{figure}
\psfig{figure=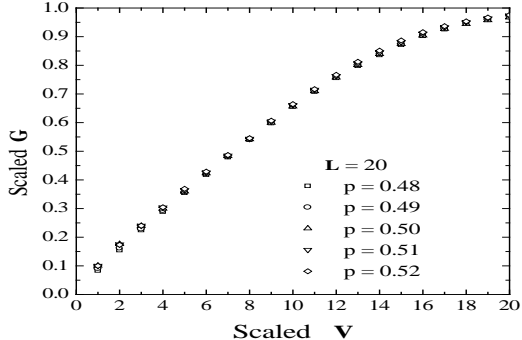,width=4cm,height=6cm}
\caption{The plot of scaled conductance against scaled voltage for various 
$p$ around $p_c$.}
\label{fig:gvsclpc}
\end{figure}

For a meaningful comparison of all the $G$-$V$ data with different $G_0$, 
$G_f$, $V_g$, {\em etc.}, we look at the scaled conductance $\tilde{G} = 
(G - G_0)/G_d$ against the scaled voltage $\tilde{V} = (V - V_g)/V_g$. 
To see if the $G$-$V$ data for various values of $p$ both below and
above $p_c$ scale, we first looked within the 
range $0.48 \le p \le 0.52$ ({\em i.e.}, very close to $p_c$), and
found that all the data do reasonably collapse.  In the
Fig.~\ref{fig:gvsclpc} we show such a plot for a $20\times 20$ system. 
This suggets the following general form for the functional behavior;
\be
\tilde{G} = f(\tilde{V}),  \label{scaling}
\ee
where $f(x)$ is a function such that $f(0)$ = 0, and $f(\infty)$ = 1 and is
otherwise quite general as long as it represents the behavior of $\tilde{G}$
very well.  Clearly the scaled function in Eq.~(\ref{gfit}) satisfy these
properties very well.      

\begin{figure}
\psfig{figure=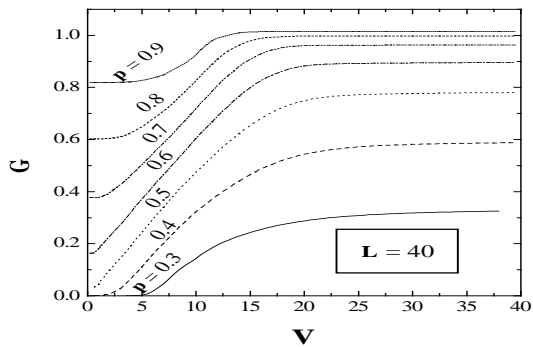,width=4cm,height=6cm}
\caption{The family of $G$-$V$ curves corresponding to the
Fig.~\ref{fig:ivset}.}
\label{fig:gvset}
\end{figure}

Here we point out that the threshold voltage
$V_g$ is the only relevant variable that enters into the 
scaling function. The other voltage scale $V_s$ which is the 
onset voltage for saturation, is seen to have no role in the above scaling
Eq.~(\ref{scaling}). For small $\Delta V=V-V_g$, {\em i.e.}, near the
onset of nonlinearity, the excess conductance $\Delta G = G - G_0$ varies with
the voltage difference ($\Delta V$) as a power-law as one may easily check
by expanding Eq.~(\ref{gfit}) in the limit ($\Delta V \rightarrow 0$): 
\be
\Delta G \sim \Delta V^{\mu\gamma} = \Delta V^{\delta}.    \label{delta}         
\ee
Thus the nonlinearity exponent $\delta$ is related to the fitting parameters
$\mu$ and $\gamma$ by $\delta=\mu\gamma$.  In our earlier work \cite{skg}
with our preliminary observations we had reported the nonlinearity
exponent $\delta$ to be close to 1 and to be independent of $p$ 
near the geometrical percolation point $p_c = 0.5$. Indeed, in most 
of the experiments an average value of the above exponent is reported for 
the data for samples close to $p_c$. 

\begin{figure}
\psfig{figure=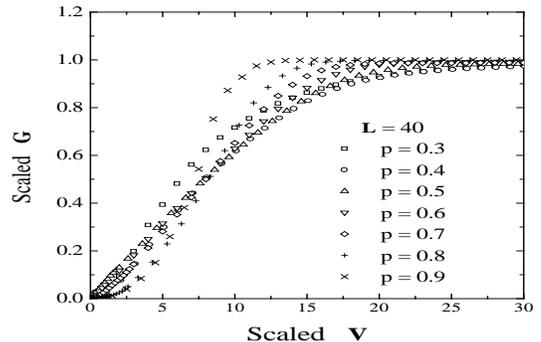,width=4cm,height=6cm}
\caption{The scaled conductance against the scaled voltage for different $p$ 
(both below and above $p_c$) corresponding to the curves in 
Fig.~\ref{fig:gvset}. The data do not collapsse at all indicating different 
exponents for different $p$'s.}
\label{fig:gvsclp}
\end{figure}

Further careful analysis of the results at widely different volume
fractions indicate that the nonlinearity exponent $\delta$ increases
significantly as we go sufficiently away from the percolation threshold
(both below and above). This becomes apparent from the shape of the $G$-$V$
curves for different volume fractions in a wide range of $p$ (from 0.3 to
0.9) in Fig.~\ref{fig:gvset} corresponding to the $I$-$V$ curves shown in
Fig.~\ref{fig:ivset}. In Fig.~\ref{fig:gvsclp} we plot the scaled 
conductance ($\tilde{G}$) against scaled voltage ($\tilde{V}$) corresponding 
to all the $G$-$V$ curves in Fig.~\ref{fig:gvset}. The scaled data for all the
curves now do not fall on top of each other indicating that all of them can 
not be described by the same fitting parameters $\mu$ and $\gamma$, even
though the form of the best fitting function $f(x)$ remains the same.
Hence the nonlinearity exponent in the power-law regime ($\Delta V \rightarrow
0^+$) for these curves of different $p$ are {\it not identical}.

\begin{figure}
\psfig{figure=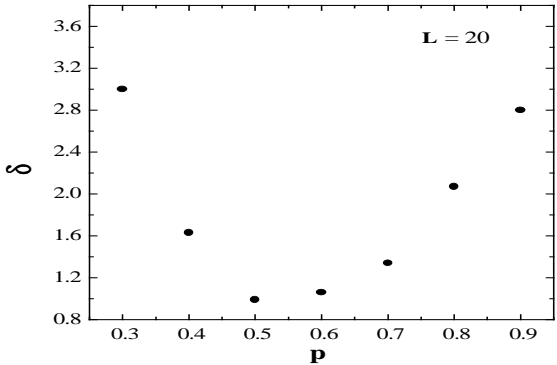,width=4cm,height=6cm}
\caption{The variation of nonlinearity exponent ($\delta$) for the $G$-$V$ 
curves with $p$ for a system of size $40\times40$.}
\label{fig:nonlinexpo}
\end{figure}

The fitting of the individual $G$-$V$ curves for squares of sizes $L$ = 20 and
40 and some data for $L$ = 60 and 80 for different $p$ in the range of 
$p$ = 0.2 to 0.9 were done by using Eq.~(\ref{gfit}). 
The Fig.~\ref{fig:nonlinexpo} demonstrates that the exponent $\delta$
for $L$ = 40 increases from a value close to 1 at $p = p_c$ to values close
to 3 and above on either side (0.3 $\le p \le $ 0.9).  We found that the value
of $\delta$ at $p_c$ lies in the range 0.97 to 1.04 for system sizes $L$ = 10
to 60. There is no systematic variation with $L$ and this indicates the
absence of any finite size dependence for the above exponent at $p_c$.
Thus within our numerical accuracy we find 
that $\delta(p_c) \cong 1.0$. It is thus clear from the 
Fig.~\ref{fig:nonlinexpo} that $\delta(p_c)$ is the minimum of the 
$p$-dependent exponent $\delta(p)$.
Note that the nonlinearity exponent $\alpha$ for the $I$-$V$ curves would be
just $\alpha = \delta + 1 \cong$ 2.0 (in this case), where 
$I \sim (V - V_g)^{\alpha}$, and for $p$ sufficiently far from $p_c$, 
$\alpha$ would also show the same concentration dependence.  In comparison,
we observe that in an experiment on 2D arrays of normal metal islands
connected by small tunnel junctions, Rimberg {\em et al.} \cite{rimberg}
found the nonlinearity exponent $\alpha$ to be 1.80 $\pm$ 0.16.
It may also be noted that Roux and Herrmann \cite{rh} also obtained from their 
simulation of the $I$-$V$ nonlinearity exponent $\alpha = 2$ in their model 
2D network with the resistors without any dilution but with random thresholds.
Clearly thus, our bond-diluted RRTN model is richer than the random threshold
model at least as far as the nonlinearity exponent is concerned.

Now we discuss the {\it crossover exponent} which is an alternative way 
of accounting for the strength of nonlinearity.  The crossover exponent
($x$) is defined from the power-law relationship $I_c \sim G_0^x$
where $I_c$ is the crossover current at which the conductance of
the system has increased from a nonzero initial value $G_0$ by a small but
arbitrarily fixed fraction $\epsilon$.  So $x$ is defined only above $p_c$. 
The value of the above exponent was calculated by Gefen {\em et. al.}, 
\cite{exp} from the experimental nonlinear response data of discontinuous
thick gold films near and above the percolation threshold (in 3D).
Their experimental measurement gives $x = 1.47 \pm 0.10$, whereas 
they argue through a model resistor network \cite{drrn} that the value of 
$x$ should be 3/2 (in 3D). The argument is based on the assumption of a power 
law dependence of conductance ($G$) which interpolates between its initial 
value $G_0$ (at $V = V_g$) and the saturation value $G_f$ ($V = V_s$).  
Note that the crossover
exponent for the carbon-wax experiment in 3D is also close to this value.
Gefen {\em et al.}~\cite{exp} found that near the threshold 
$\delta = t/\nu$. Thus one expects that close to $p_c$, $\delta = 1.3/1.33
\cong$ 0.97 and the nonlinearity exponent for $I$-$V$ characteristic is thus 
$\alpha = \delta+1 \cong 1.97$ in 2D. It will be noted that close to $p_c$, 
the nonlinearity exponent for our model in 2D is very close to it. Further,
since the exponent $\delta$ lies between about 3 and 1 for $p$ between 0.3
and 0.9, the crossover exponent in our model can vary between 1.3 and 2 in
the same dilution range in 2D.

Apart from calculating the nonlinearity exponent as discussed previously, 
one may also measure or calculate the difference between the conductance
($G$) in the two {\it asymptotic linear regimes}.
The two limiting conductances $G_0$ and $G_f$ are already defined in 
two asymptotic linear regimes namely, $V \rightarrow 0$ and 
$V \rightarrow \infty$ respectively.  Whereas the nonlinearity exponent
($\delta$ or $\alpha$) acts as a measure of the {\it initial nonlinearity}
near the threshold $V_g$, the difference in conductance 
$G_d = G_f - G_0$ serves as a measure of the {\it overall nonlinearity}.
We have the following criteria which may be easily checked.
\begin{itemize}
\item In the range $p \le p_{ct}$:~ $G_0 = 0$ and $G_f = 0$, so $G_d = 0$,
\item In the range $p_{ct} < p \le p_c$:~ $G_0 = 0$ and $G_f \ne 0$, so 
$G_d = G_f$,
\item In the range $p_c < p < 1$:~ $G_0 > 0$ and $G_d = G_f - G_0$. 
\end{itemize}

\begin{figure}
\psfig{figure=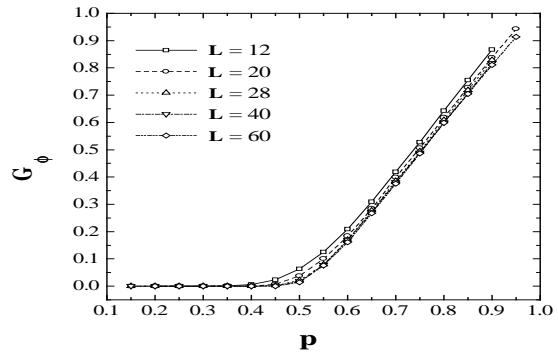,width=4cm,height=6cm}
\caption{The behavior of $G_0$ against $p$. 
The system sizes ($L$) are indicated in the figure.}
\label{fig:g0p}
\end{figure}

\begin{figure}
\psfig{figure=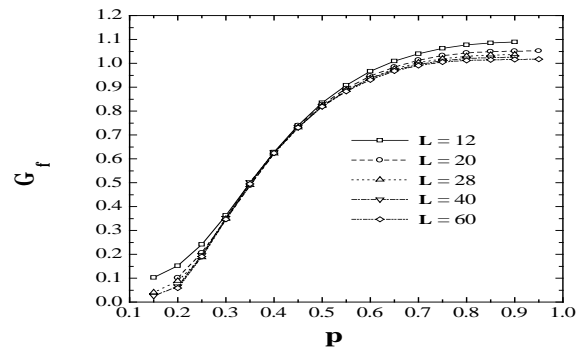,width=4cm,height=6cm}
\caption{The behavior of $G_f$ against $p$.
The system sizes ($L$) are indicated in the figure.}
\label{fig:gfp}
\end{figure}

\begin{figure}
\psfig{figure=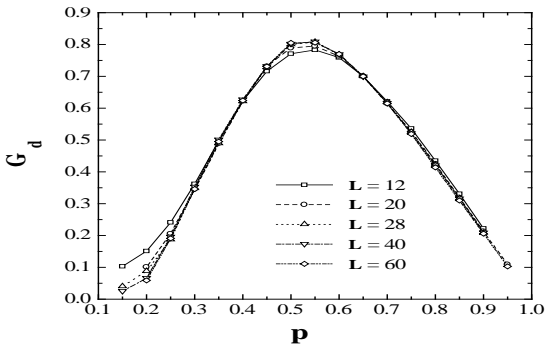,width=4cm,height=6cm}
\caption{The behavior of $G_d$ against $p$ for different system sizes $L$ 
as indicated in the figure.}
\label{fig:gdp}
\end{figure}

We have shown the variation of $G_0$ and $G_f$ against $p$ in 
Fig.~\ref{fig:g0p} and in Fig.~\ref{fig:gfp} respectively for a
square network of size $L$ = 12, 20, 28, 40 and 60.  The averages are done
over 100 to 1000 configurations in each case. The interesting point to
note is that as a function of $p$, $G_f$ tends to be flat ($p$-independent) as
$p \rightarrow 1$ but $G_0$ seems to be increasing sharply.  Further, one
may note that while $G_0$ behaves in a power-law fashion around $p_c = 0.5$, 
$G_f$ does the same around $p_{ct} \cong 0.181$. Next, the difference in 
conductance $G_d$ is plotted against $p$ in Fig.~\ref{fig:gdp}. 
The peak in this curve appears at around $p = p_c$. This indicates that the 
overall nonlinearity is maximum near the geometrical percolation threshold.
There does not seem to be any strong finite-size effects in the $G_d$-$p$
graph except close to $p_{ct}$.

\begin{figure}
\psfig{figure=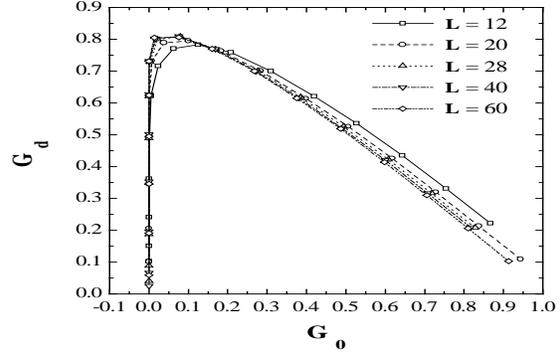,width=4cm,height=6cm}
\caption{The $G_d$ against $G_0$ for different $L$ as indicated in the 
figure.}
\label{fig:gdg0}
\end{figure}

\begin{figure}
\psfig{figure=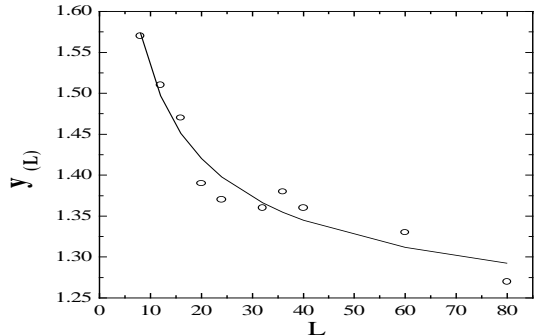,width=4cm,height=6cm}
\caption{The finite size behavior of $y(L)$, for different $L$ data fitted
with $y(L) = y(\infty) + B/L^r$ where $y(\infty)$ = 1.1, $B$ = 1.2, $r$ = 0.5}
\label{fig:ylfsize}
\end{figure}

The next important question is how the difference in conductance $G_d$ is
related to the initial conductance $G_0$. The variation of $G_d$ against
$G_0$ is shown in the Fig.~\ref{fig:gdg0} for the whole range of $p$
(0 to 1). As expected, the peak appears at around $p = p_{ct}$, and a strong
finite size effect may be noted in the $G_d$-$G_0$ graph.  One may also
observe that in the thermodynamic limit, $G_d = 0$ for $p \le p_{ct}$.
Now we would like to examine the above relationship, {\em i.e.}, $G_d$ against 
$G_0$ in the interval $p_{ct} < p < 1$.  We do fit the data with the function 
$G_d = A - BG_0^y(L)$, where $A$ and $B$ are constants and the exponent 
$y(L) > 0$.  The data were fitted for different system sizes from $L$ = 8 to
80.  In the Fig.~\ref{fig:ylfsize}
we have shown $y(L)$ against $L$ to show the finite size behavior and find 
that $y(\infty) \cong $ 1.1 which is close to 1.  Thus we may conclude
that $G_d$ is almost linearly dependent on $G_0$ in the limit of
$L \rightarrow \infty$ which means that $G_f$ is also linearly dependent on 
$G_0$. This in turn supports the idea of identical scaling relationship for two 
saturation conductances $G_0$ and $G_f$ around the respective thresholds 
($p_c$ and $p_{ct}$). This is consistent with the fact that the system is in 
the same {\it universality class} in both the limits ({\em i.e.}, 
$G_f \simeq (p-p_{ct})^t$).

\subsection{Comparison with some EMA results}

\begin{figure}
\psfig{figure=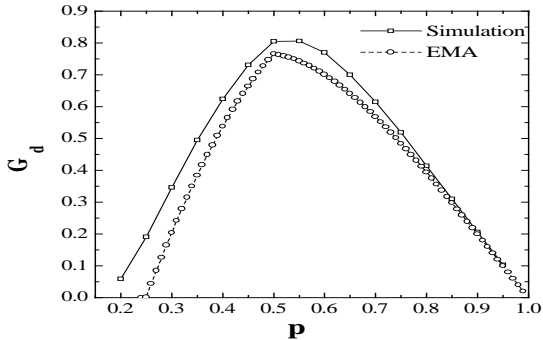,width=4cm,height=6cm}
\caption{The variation of $G_d$ against $p$, obtained by the simulation and 
the EMA, are plotted for comparison.}
\label{fig:gdpema}
\end{figure}

\begin{figure}
\psfig{figure=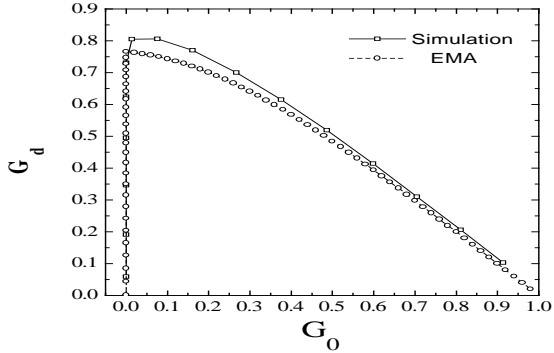,width=4cm,height=6cm}
\caption{The variation of $G_d$ against $G_0$, obtained by the simulation and 
the EMA, are plotted for comparison.}
\label{fig:gdg0ema}
\end{figure}

For the sake of comparison we plot the $G_d$ against $p$ and  $G_d$ against
$G_0$ curves with the corresponding results obtained by EMA for our model
(in 2D). In Fig.~\ref{fig:gdpema} the $G_d$ against $p$ is plotted for
both the simulation result and the EMA result. Both are shown for our model
network in 2D, where the simulation result plotted is for $L = 60$. 
In Fig.~\ref{fig:gdg0ema} we plot $G_d$ against $G_0$ again for the 
simulation and the EMA result. It is apparent from the above two figures 
that these results are close. 

The effective conductance of the network $G_e = G_0$ when all the 
tunneling bonds behave as perfect insulators (the usual limit of binary 
mixture of conductors and insulators, $p_c = 1/2$) and the effective 
conductance $G_e = G_f$ when all the tunneling bonds (or resistors) behave 
as the ohmic conductors (Eq.~(\ref{sol})).
Clearly, if one puts $P_t=0$ in Eq.~(\ref{sol}), one finds that
\be
G_0 = {dP_o - 1 \over (d-1)}.
\ee 
The behavior of $G_d$ against $p$ is shown in Fig.~\ref{fig:gdpema} for 
2D.  $G_d$ has a maximum at $p_c$ (= 1/2 in 2D) as 
it should be because the system with the $o$- bonds only is most tenuous at 
the geometrical percolation threshold and the measure of nonlinearity clearly 
should be maximum there (with the largest concentration of the $t$-bonds). 
The plot of $G_d$ against $G_0$ is also shown in Fig.~\ref{fig:gdg0ema}. 
In the same plot we have also shown the simulation result for comaprison and
one may note that the quantity $G_d$ is maximum at $p = p_c$.  The EMA result
is quite close to the simulation result and the match becomes better as
one goes farther and farther away from $p_{ct}$.

\section{Summary and Comments}

In this paper, we have been mainly concerned with the nonlinear response
chracteristics, the relevant scaling and related exponents in composites
or granular metallic systems where transport due to charge tunneling plays
an important role.
One may note that our correlated percolation model and the corresponding
RRTN network can capture the essential features related to nonlinearity 
quite well.  Before ending we would like to make some comments and
discussions for the sake of completeness. 
The existence of a well-defined voltage threshold ($V_g$) close to which a
power-law regime appears indicates that this threshold (breakdown) voltage is 
a {\it critical point}. Similar threshold behavior has been observed in 
pinned charge-density-wave (CDW) conductors, superconductors with pinned 
vortices (type-II), {\em etc}. Indeed it has been argued long ago that
wherever there exists some sort of threshold of force for motion to occur,
the threshold actually corresponds to a {\it dynamic critical point} 
\cite{dcrit} for the driven dynamical system.  Disorder in such systems is
known to give rise to `pinning' or inhibition to transport upto a critical 
value of the force. 
Clearly in our percolation model, the threshold $V_g$ acts as a dynamical
critical point for the systems with volume fractions in the range
$p_{ct} < p < p_c$.  For such systems the field corresponding to the
threshold voltage $V_g$ is called the {\it dielectric breakdown field} and 
is pretty well-studied in the DRRN \cite{drrn} type models. We have focussed 
on the threshold as a dynamical critical point in our discrete model, and 
calculated the breakdown exponent for our RRTN model elsewhere \cite{brdn}. 

As we pointed out above (see also Ref.~\cite{skg}) that
the mechanism of nonlinearity is essentially the same both below and above the 
system threshold. Below the system thresold, there is no system spanning 
cluster. So the transport is identified as {\it intercluster} tunneling or 
hopping across dangling bonds or gaps. Above the threshold 
there are both intercluster and intracluster tunneling. But {\it intracluster}
tunneling mechanism certainly dominates. The nearest neighbor gaps are 
everywhere; both inside the smaller isolated clusters as well as in the 
system spanning cluster. So the tunneling mechanism is operative both below
and above $p_c$ (in the interval $p_{ct} < p < 1$) giving rise to nonlinear
regime in the response.

From our analysis of current-voltage ($I$-$V$) data we understand that one does
not do justice by fitting only the $I$-$V$ curve and finding out the 
nonlinearity exponent therefrom since that fitting is not robust. One may easily 
get tempted to fit the nonlinear regime of a $I$-$V$ curve in general through 
an $n$-th degree polynomial function. A reasonable choice \cite{rkc} would be to 
fit with the law: $I = G_1V + G_2V^3$, assuming that the leading nonlinear term 
is cubic (ignoring other higher order terms). Even ignoring the fact that
the nonlinearity exponent away from $p_c$ are significantly different from
integer values, we note that this type of analysis in this form may lead to
confusion. The determination of the exponents and the coefficients may become 
arbitrary. Arbitrary selection of the range of $I$-$V$ data for this purpose 
and the fitting of that would account for this arbitrariness. One may not know 
upto what voltage scale (in the nonlinear regime) one should fit the data and 
hence one may get a set of different answers for the values of the exponents. 
A typical $I$-$V$ curve, for the kind of systems we address, can in general be 
fitted by a simple power-law: $I \sim (V - V_g)^{\alpha}$, at least around the 
threshold voltage $V_g$. 
But this also may not provide unambiguous answer most of the time.
One may note that the exponent for such power-law does change continuously
as the applied voltage $V$ is increased from $V_g$ and the $I$-$V$ curve 
may ultimately approach another linear regime. In fact, the selection of the
range of data and its fitting had been at the root of many confusing results
as found in the literature. Instead our prescription of fitting of the 
$G$-$V$ data for the entire nonlinear regime along with the saturation, as 
presented, is found to be most satisfactory.
>From such a fitting (Eq.~(\ref{gfit})) of the $G$-$V$ data, one can obtain 
the desired power-law and find the exponent. The $G$-$V$ curves for the 
type of composites we focuss our attention on, may all be generically fitted 
by a function like $G(V) = G_0 + G_df(\Delta V)$, where
$f(\Delta V)$ is a function that behaves as $\Delta V^{\delta}$ for small 
$\Delta V$, where $\Delta V = V - V_g$. $f(\Delta V) \rightarrow 0$ as 
$\Delta V \rightarrow 0$ and $f(\Delta V) \rightarrow 1$ as 
$\Delta V \rightarrow \infty$ (or an appropriately large value for a given
finite system). 

Next we point out a few observations on some recent experiments in the
literature which are found to be not in full agreement with our simulation
results. In a recent experiment on the carbon-wax mixture \cite{nan} it is 
claimed that the conductivity exponent ($t$) to be different in the two extreme
limits namely, $\Delta V \rightarrow 0$ and $\Delta V \rightarrow \infty$.
More explicitly and in the perspective of the foregoing discussion, 
$G_0 \sim \Delta p^t$ and $G_f \sim \Delta p^{t^{\prime}}$, where $\Delta p =
(p-p_c)$.  Note that the $\Delta p$ for the scaling of $G_f$ should have been
$(p-p_{ct})$ as we have discussed before. In any case, it is claimed 
that $t^{\prime} \ne t$. More specifically, $t^{\prime} = ct$, where
$c \cong 0.76$ (in 3D) as reported \cite{nan}. This in turn is related to the 
question of universality class, claiming that the system goes over to different 
universality class in the saturation limit ($V \rightarrow \infty$) (has been 
termed as altered percolating state in Ref.~\cite{nan}).
However, ws we discuss in section~\ref{nonlinexpo} (we have also claimed 
earlier \cite{kgs}) that the system seems to be in the 
same universality class in both the limits whereby we claim that 
$t = t^{\prime}$ within the numerical accuracy. 
Our calculation of course is in 2D. But the essential physics should remain
the same as we go over to 3D.

We have seen that the nonlinearity exponent, $\alpha$ ($= \delta + 1$) 
varies significantly with the volume fraction ($p$) of the conducting component,
the minimum value of the exponent being around 2.0 at $p = p_c$.
Chen and Johnson, in their experiment on $Ag$-$KCl$ \cite{chen}, found the
above nonlinearity exponent to vary with $p$. In particular they found that 
for a sample very close to percolation threshold, the nonlinearity 
exponent is as large as 20. In another early and a very prominent experiment, 
nonlinear response in $ZnO$ varistors has been addressed by Mahan 
{\em et al.} \cite{mahan}. The observed $I$-$V$ characteristics in the $ZnO$ 
varistors are often empirically described by the power-law relation:
\be
I = kV^{\alpha}, 
\ee
where the parameter $\alpha$ ($> 1$) is the measure of nonlinearity.
A theory had been developed by Mahan {\em et al.} to predict the coefficient of
nonlinearity ($\alpha$) as high as 50 or even 100.  
Such a high value of $\alpha$ of course indicates an exponential relationship 
rather than a power-law. Canessa and Nguyen \cite{can} had attempted a
computer simulation study of the varistors considering a binary mixture
model to understand the very high value of $\alpha$. 

\begin{figure}[p]
\psfig{figure=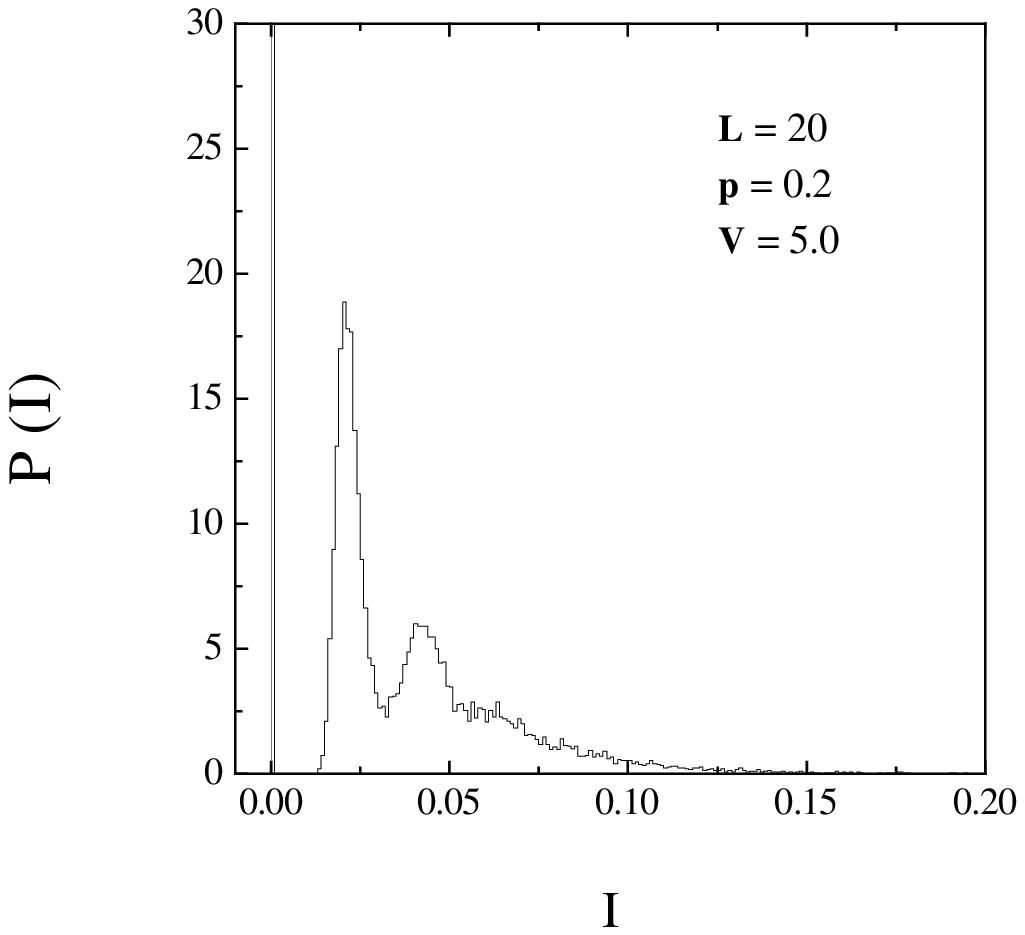,width=4cm,height=6cm}
\caption{The distribution of current ($I$) in the 2D square network of
size $L$ = 20 for $p$ = 0.2 and an applied voltage $V$ = 5.0. This 
distribution is non-self-averaging. 6$\times10^4$ configurations were 
used for this purpose.} 
\label{fig:distv5}
\end{figure}

\begin{figure}[p]
\psfig{figure=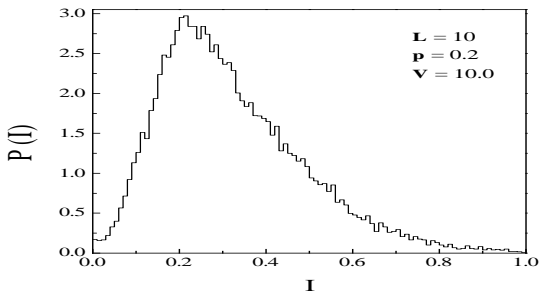,width=4cm,height=6cm}
\caption{The distribution of current ($I$) in the 2D square network of
size $L$ = 20 for $p$ = 0.2 and an applied voltage $V$ = 10.0.
This distribution is self-averaging. 2$\times10^4$ configurations were 
used for this purpose.} 
\label{fig:distv10}
\end{figure}

Intrigued by these results we ran some test for a 
square lattice with ($L$=20 and 40) $p$=0.2, very close to our $p_{ct}$. 
We obtained in our preliminary analysis by fitting $G$ with voltage $V$ that
$\alpha > 20$  for $L = 20$ and $\alpha \gg 1$ (diverging for ever) for
$L = 40$.
This was very suspicious and by zooming into the fitting close to
the threshold ($V_g$), we found the fitting to be extremely bad. The 
other functions listed before (including the double-exponential) did not
give any better fitting either. But this is the region which gives the
initial power-law exponent. To get the fit better in this region, we fitted
the logarithmic conductance (ln~$\Delta G$), keeping our fitting function the
same Eq.~(\ref{gfit}) as before. Then we found the exponent $\delta$ to be of
the order of 3 (the Fig.~\ref{fig:nonlinexpo} was done by taking care of this 
fitting problem). Thus power-law description still seems to hold but there
seems to be a different problem close to $p_{ct}$. Since the fitting was
still not as good as at higher $p \ge$ 0.4, we wanted to check if the
averaging process itself is at flaw at very low volume fractions. Hence
we look at the distribution of the current ($I$) for different realizations
of the sample of size $L$ = 20 and $p$ = 0.2 for two different voltages.
In Fig.\ref{fig:distv5}, the histogram is shown for $V$ = 5, and in
Fig.\ref{fig:distv10}, the same is shown for $V$ = 10. The distribution for 
$V$=10 is reasonably well behaved, but the one for $V$=5 has an isolated
delta-function-like peak at zero conductance apart from two other broader 
peaks at higher conductances. We did also look at the relative value of the 
variance (relative to the mean squared) defined as
\be
RV = {(<I^2> - <I>^2) \over <I>^2}
\ee
If this value is less than 1, averaging is alright, but if it is much
larger than 1, the self-averaging property of the distribution does not
work. We have tested that $RV$ is about 3.3 for the set of data for $V$ = 5 
and it is about 0.3 for $V$=10.
Thus the low $p$ samples are non-self-averaging at low voltages but tend
to be more self-averaging at higher voltages. This property was also
reported for quantum systems by Lenstra and Smokers \cite{lens}, and
thus our semi-classical model of percolation indeed seems to capture 
some non-classical (quantum) behavior at low $p$ as expected. 

The connection between the above exponent $\delta$ and the crossover exponent
($x$) is the following. The crossover voltage $V_c$ is related to the 
initial nonzero conductance $G_0$ by $V_c \sim G_0^{x-1}$. But from 
Eq.~(\ref{delta}) we know that the excess conductance $\Delta G(V) \cong 
\Delta V^{\delta}$ for small $\Delta V$. Choosing $\Delta G \cong \epsilon G_0$
for an arbitrary but small
$\epsilon$ and if one is close to percolation threshold, one has 
$V_c \sim G_0^{1/\delta}$ which implies that $\delta = 1/(x-1)$.
Thus, as the value of nonlinearity exponent is dependent on $p$, the crossover
exponent ($x$) should also be dependent on $p$. The values of the crossover
exponent $x$ so far reported in the literature are for the experimental data for the samples 
close to but above the threshold. But we may conclude that this exponent should 
also show change with $p$ if $p$ is away from $p_c$. In fact, such an
example may be mentioned here.
The crossover exponent is found to be widely different for a set of samples
as measured recently \cite{bar}. The reason for this is yet to be understood.
But if the samples taken for measurements have
widely different volume fractions ($p$) of conducting components, then 
nonlinearity exponent and thus the crossover exponent for them could be widely
different. 

Now it may be noted that if $\delta > 1$ or in other words $\mu\gamma > 1$ in 
the Eq.~(\ref{gfit}), the first derivative of $G$ with respect to
$V$, {\em i.e.}, $dG/dV$ would attain a maximum value at the inflexion point
(at some voltage in the nonlinear regime).  
We fitted a set of $G$-$V$ and the corresponding $dG/dV$-$V$ data (see 
Ref.~\cite{thesis}), taken from the experimental observations on a carbon-wax 
composite system \cite{kkb} to show the appearance 
of a peak in the $dG/dV$-$V$ curve. To indicate how good or bad our fitting
equation is, we fit the above data set for $G$-$V$ characteristics by the
Eq.~(\ref{gfit}). The fitted line is seen to match the experimental data
very well with the above function. As a further test of the fitting, we took
the parameters of the $G$-$V$-fit, and used them (the functional form) to
obtain $dG/dV$ as a function of $V$.
The agreement with the experimental values for $dG/dV$ should be
considered rather good given that $G$ and $dG/dV$ are independent measurements
and that higher the harmonic more error-prone is its value.  Incidentally,
the nonlinearity exponent $\delta$ for this 3D experimental data fitted by
our method comes out to be 1.74, where the crossover exponent measurement
on the same sample would give an $\delta$-value of about 2. 

\section{acknowledgement}

We thank K.K. Bardhan, R.K. Chakrabarty and U. Nandy
for allowing us to go through their experimental data and for the related 
discussions. We are also thankful to B.K. Chakrabarti, S. Roux and 
H.J. Herrmann for many useful discussions.


\begin{thebibliography}{9}
\bibitem{note2} Even in this semi-classical sense, these
barriers are still fundamentally different from the Ohmic resistors since
they contribute to the resistance but not to heat dissipation.
\bibitem{note3} It may be noted that
if we include tunneling between next nearest neighbors also, the
effective percolation threshold would go down even further.
For simplicity, in this work we restrict ourselves to nearest neighbor 
tunneling only.
\end{thebibliography}
\end{document}